\documentclass[a4paper,11pt]{article}
\usepackage{amsmath}
\usepackage{amssymb}

 \newread\testifexists
\def\GetIfExists #1 {\immediate\openin\testifexists=#1
    \ifeof\testifexists\immediate\closein\testifexists\else
    \immediate\closein\testifexists\input #1\fi}

\immediate\openin\testifexists=e
    \ifeof\testifexists\immediate\closein\testifexists\else
    \immediate\closein\testifexists\input e\fipsf

\begin{document}

\title{ $B_s \to \rho(\omega) K^{\ast}$ with perturbative QCD approach}
\author{Jin Zhu \footnote {zhujin@mail.ihep.ac.cn} \,, Yue-Long  Shen \footnote {shenyl@mail.ihep.ac.cn}
      \\
{\it \small   Institute of High Energy Physics, CAS, P.O.Box
918(4)   Beijing 100049,   China}\\
 {\it \small   Graduate School of Chinese Academy of
Science, Beijing 100049, China }\\
Cai-Dian L\"u\,\\
{\it \small  CCAST (World Laboratory), P.O. Box 8730,
   Beijing 100080,  China}\\
{\it \small   Institute of High Energy Physics, CAS, P.O.Box
918(4)   Beijing 100049,  China\footnote {Mailing address}}\\}
\maketitle

\begin{abstract}
 $B_s \to \rho(\omega)
K^{\ast}$ are useful to determine the $B_s$ distribution
amplitude, as well as constrain the CKM phase angle $\alpha$. We
study these decays within the Perturbative QCD  approach(PQCD). In
this approach, we calculate factorizable, non-factorizable, as
well as annihilation diagrams. We find the branching ratio for
$B_s \to \rho^+ K^{*-}$ is at the order of $10^{-5}$, and there's
large direct CP violation in $B_s(\bar B_s) \to \rho^0 \bar
K^{*0}(\omega K^{*0})$. Our predictions are consistent with those
 from other methods and current experiments.
\end{abstract}

\section{Introduction}
\hspace*{\parindent} Exclusive nonleptonic B decays have provided
a fertile field to investigate the CP violation and search for new
physics. The hadronic matrix elements of the effective operators
play a key role in the study of B meson decays, but it is
difficult to calculate them precisely due to the long distance QCD
dynamics. The factorization approach (FA) \cite{bsw,ali} based on
the color transparency mechanism has been applied to many decay
modes, and it works well in many channels. But it suffers from
some problems such as infrared-cutoff and scale dependence. To
solve these problems and make more accurate predictions, the
perturbative QCD approach (PQCD) \cite{brodsky,hnli,keum,pipi},
the QCD improved factorization (QCDF) \cite{bbns,beneke} as well
as the Soft-collinear effective theory (SCET) \cite{scet} have
been developed in the recent years.

PQCD is based on $k_T$ factorization theorem \cite{kt1,kt2,kt3}.
The decay amplitude is factorized into the convolution of the
mesons' light-cone wave functions (see $Appendix\ A$), the hard
scattering kernels and the Wilson coefficients, which stand for
the soft, hard and harder dynamics respectively. The transverse
momentum is introduced so that the endpoint singularity which will
break the collinear factorization is regulated and the large
double logarithm term appears after the integration on the
transverse momentum, which is then resummed into the Sudakov form
factor. The formalism can be written as:
\begin{eqnarray}
{\cal M}\sim\nonumber &&\int
dx_1dx_2dx_3b_1db_1b_2db_2b_3db_3Tr[C(t)\Phi_B(x_1,b_1)\Phi_{K^*}(x_2,b_2)
\Phi_{\rho}(x_3,b_3)\\ &&H(x_i,b_i,t)S_t(x_i)e^{-S(t)}],
\end{eqnarray}
where the $b_i$ is the conjugate space coordinate of the
transverse momentum, it denotes the transverse interval of the
meson. $t$ is the energy scale in hard function $H$. The jet
function $S_t(x_i)$ comes from the summation of the double
logarithms $\ln^2x_i$ near the endpoint, called threshold
resummation \cite{kt1,threshold}. The factorization theorem
guarantees the infrared safety and the gauge invariance of the
hard kernel and has been proved to all order of $\alpha_s$
\cite{proof}.

Many hadronic two body $B$ decays have been studied in  PQCD
approach \cite{keum,pipi,phikstar,pqcdexample}.  Most predictions
are consistent with the current experiments. The $B_s$ decays are
important to extract CKM phase angles and study the CP violation.
As $B_s$ meson is not in the energy scale of the high luminosity B
factories SLAC and KEK, it is more difficult to be produced and
measured now. We can study the $B_s$ decays more precisely in the
very near future with  the increase of luminosity at TEVATRON and
the upcoming Large Hadron Collider (LHC).

  $B_s$ meson is different from B meson due to the heavier
strange quark (compare to $u$, $d$ quark) which induces the SU(3)
symmetry-breaking effect. This effect is considered to be small
and the distribution amplitude of $B_s$ meson(given in the
following formula) should be similar to that of the B meson,
\begin{eqnarray}
\phi_{B_s}(x)=N_{B_s}x^2(1-x)^2\exp\left[-\frac{1}{2}(\frac{xM_{B_s}}
{\omega_{B_s}})^2-\frac{\omega^2_{B_s}b^2}{2}\right].\label{bwave}
\end{eqnarray}
 The upper limit of the $B_s \to \pi K$  branching
ratio is $7.5\times 10^{-6}$ \cite{bspikex}, which  constrain the
parameter $\omega_{B_s}$ to a lower limit of about 0.5
\cite{bspik}. Moreover, in order to fit the branching ratio
measured in the $B_s \to \phi\phi$ decay \cite{bsphiphiex}, we
constrain $\omega_{B_s}$ to about 0.55 \cite{bsphiphi}, then we
can see that the SU(3) symmetry-breaking is not negligible. Here
we integrate out the variable $b$ and show the distribution
amplitude of $B$ and $B_s$ meson in Fig.1. We can see that the
peak point of the  curve of $B_s$ meson's distribution amplitude
prefers a larger $x$ ($x$ denotes the momentum fraction carried by
the light quark) region comparing to the B meson. This is
consistent with the fact that the s quark  much heavier than the d
(u) quark, should carry more momentum. Later in this paper, we
will see that the branching ratios of $B_s$ decays are very
sensitive to this parameter. If measured by experiments, radiative
leptonic decays of $B_s$ meson can provide information of this
parameter \cite{nunu}.

\begin{figure}[tbh]
\begin{center}
\epsfxsize=4.0in\leavevmode\epsfbox{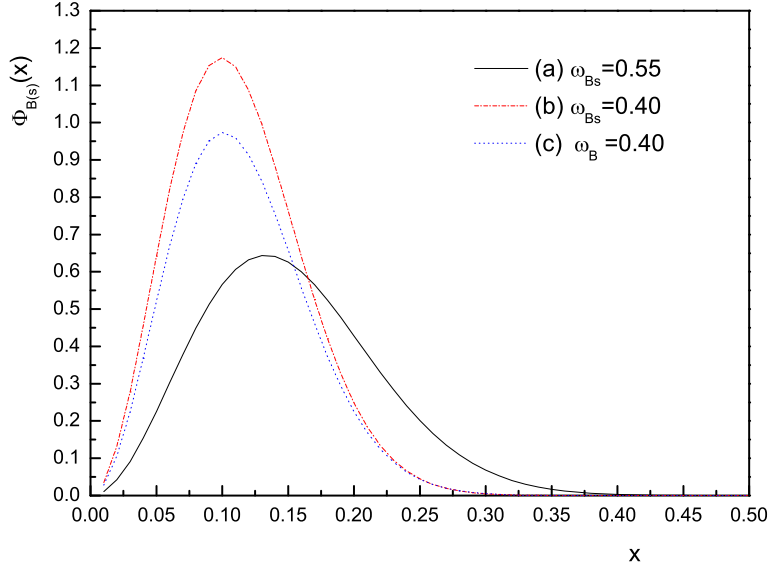}
\end{center}
\caption{{\protect\footnotesize $B_{(s)}$ distribution amplitude
as a function of light quark momentum fraction $x$ (The wave
function comes to zero when $x \in \{0.5,1.0\}$ so we don't show
it here)}}
\end{figure}

In this paper, we  study $B_s \to \rho(\omega) K^*$ decays in the
PQCD approach. Hopefully the branching ratio is not too small and
can be detected by the TEVATRON or LHCb experiments, then it may
allow us to determine the $B_s$ distribution amplitude and SU(3)
breaking effects with much more precision. Moreover, we can also
constrain $\alpha$ with fewer pollution from this channel.

\section{Calculation and Numerical analysis}

\hspace*{\parindent}We use the effective Hamiltonian for the
process $B_s \to \rho(\omega) K^*$ given by \cite{Buchalla}
\begin{eqnarray}
{\cal
H}_{eff}=\frac{G_F}{\sqrt{2}}\left\{V_u\left[C_1(\mu)O_1(\mu)+C_2(\mu)O_2(\mu)\right]
-V_t\sum_{i=3}^{10}C_i(\mu)O_i^{(q)}(\mu)\right\},
\end{eqnarray}
where  $V_u=V_{ud}^*V_{ub}$, $V_t=V_{td}^*V_{tb}$, $C_i(\mu)$ are
the Wilson coefficients, and the operators are
\begin{eqnarray}
\nonumber &&O_1=(\bar{d}_iq_j)_{V-A}(\bar{q}_jb_i)_{V-A},\,
O_2=(\bar{d}_iq_i)_{V-A}(\bar{q}_jb_j)_{V-A}, \\
\nonumber
&&O_3=(\bar{d}_ib_i)_{V-A}\sum\limits_q(\bar{q}_jq_j)_{V-A},\,
O_4=(\bar{d}_ib_j)_{V-A}\sum\limits_q(\bar{q}_jq_i)_{V-A}, \\
\nonumber &&
O_5=(\bar{d}_ib_i)_{V-A}\sum\limits_q(\bar{q}_jq_j)_{V+A},\,
 O_6=(\bar{d}_ib_j)_{V-A}\sum\limits_q(\bar{q}_jq_i)_{V-A},\\ \nonumber
 &&O_7=\frac{3}{2}(\bar{d}_ib_i)_{V-A}\sum\limits_qe_q(\bar{q}_jq_j)_{V+A},\,
 O_8=\frac{3}{2}(\bar{d}_ib_j)_{V-A}\sum\limits_qe_q(\bar{q}_jq_i)_{V+A},
 \\ &&
 O_9=\frac{3}{2}(\bar{d}_ib_i)_{V-A}\sum\limits_qe_q(\bar{q}_jq_j)_{V-A},\,
 O_{10}=\frac{3}{2}(\bar{d}_ib_j)_{V-A}\sum\limits_qe_q(\bar{q}_jq_i)_{V-A}.
 \end{eqnarray}
Here i and j stand for $SU(3)$ color indices.

 The
decay width for these channels is :
\begin{eqnarray}
\Gamma=\frac{G_F^2|{\bf p}|}{16\pi M_B^2}\sum\limits_{\sigma
=L,T}{\cal M}^{\sigma\dag}{\cal M}^{\sigma}
\end{eqnarray}
where ${\bf p}$ is the  3-momentum of the final state mesons,
$|{\bf p}|=\frac{M_B}{2}(1-r^2_{K^*}-r^2_{\rho(\omega)})$, and
$r_{K^*(\rho,\omega)}=m_{K^*(\rho,\omega)}/m_{B_s}$. ${\cal
M}^{\sigma}$ is the decay amplitude , which will be calculated
later in PQCD approach. The subscript $\sigma$ denotes the
helicity states of the two vector mesons with the longitudinal
(transverse) components L(T). According to Lorentz structure
analysis, the amplitude can be decomposed into:
\begin{eqnarray}
{\cal M}^{\sigma}=M_{B_s}^2{\cal M}_L+M_{B_s}^2{\cal
M}_N\epsilon^{\ast}_2(\sigma=T)\cdot \epsilon^{\ast}_3(\sigma=T)+
i{\cal
M}_T\epsilon_{\mu\nu\rho\sigma}\epsilon_2^{\mu\ast}\epsilon_3^{\nu\ast}P_2^{\rho}P_3^{\sigma}.
\end{eqnarray}
We can define the longitudinal $H_0$, transverse $H_{\pm}$
helicity amplitudes
\begin{equation}
H_0=M^2_{B_s}{\cal M}_L,\,\,H_{\pm}=M^2_{B_s}{\cal M}_N\mp
M_{K^*}^2\sqrt{r^{\prime 2}-1}{\cal M}_T,
\end{equation}
where $r^{\prime}=\frac{P_2\cdot P_3}{M_{K^*}M_{\rho(\omega)}}$.
They satisfy the relation
\begin{equation}
\sum\limits_{\sigma =L,R}{\cal M}^{\sigma\dag}{\cal
M}^{\sigma}=|H_0|^2+|H_+|^2+|H_-|^2.
\end{equation}

The leading order diagrams in PQCD approach are shown in Fig.2.
The amplitudes for $B_s \to \rho {K}^{\ast0}$ and $\bar{B}_s\to
\rho\bar{K}^{\ast0}$are written as
\begin{eqnarray}
{\cal M}_{H}&=&V_u^*T_H-V_t^{*}P_H,\label{m}
\\
\bar {\cal M}_{H}&=&V_uT_H-V_tP_H,\label{barm}
\end{eqnarray}
respectively, where the subscript $H$ denote different helicity
amplitudes $L,N$ and $T$, and $T_H$ and $P_H$ are the amplitudes
 from tree and penguin diagrams respectively. The detailed formulae
of $T_H$ and $P_H$ are similar to those in $B \to K^*K^*$
\cite{kstarkstar} and $B \to \phi K^*$ \cite{phikstar}, so we will
not show them here.

\begin{figure}[tbh]
\begin{center}
\epsfxsize=3.0in\leavevmode\epsfbox{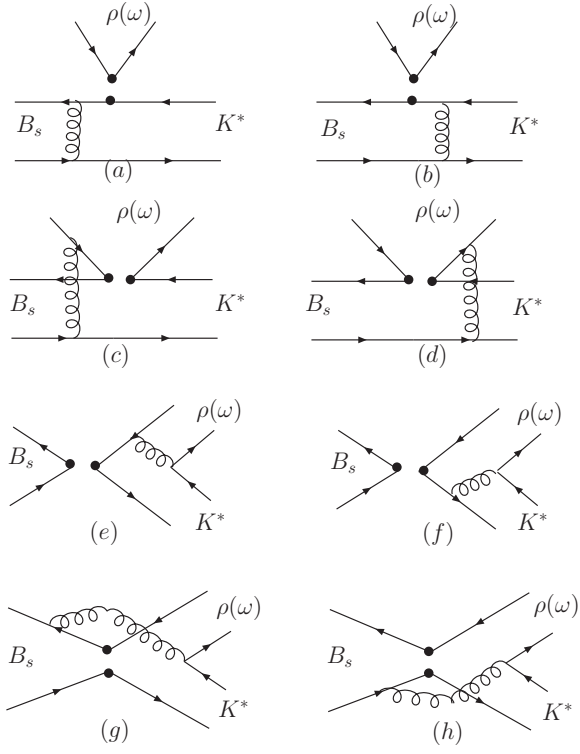}
\end{center}
\caption{{\protect\footnotesize Leading order Feynman diagrams for
$B_s\rightarrow \rho (\omega) K^{*}$}}
\end{figure}

 The parameters used in our calculations are:
the Fermi coupling constant $G_F=1.16639\times10^{-5}GeV^{-2}$,
the meson masses $M_{B_s}=5.37GeV,\ M_{K^*}=0.89GeV,\
M_{\rho(\omega)}=0.77GeV$ \cite{particle}, the decay constants
$f_{K^*}=0.217GeV,\ f_{K^*}^T=0.16GeV,\ f_{\rho}=0.205GeV,\
f_{\rho}^T=0.155GeV,\ f_{\omega}=0.195GeV,\ f_{\omega}^T=0.14GeV$
\cite{fkstar}, the central value of the CKM matrix elements
$\alpha=95^\circ$, $|V_{td}|=0.0075,\ |V_{tb}|=0.9992,\
|V_{ud}|=0.9745$ \cite{particle}, $|V_{ub}|=0.0047$
\cite{baverage} and the $B_s$ meson lifetime $\tau_{B_s}=1.461ps$
\cite{particle}.

 \begin{table}
\caption{ Branching ratio, polarization fraction and direct CP
asymmetry dependence on $\omega_{B_s}$}
\begin{tabular}{|c|c|c|c|c|c|c|}
 \hline \hline
Channel& $\omega_{B_s}$& BR$(10^{-6})$ & $|A_0|^2$ & $|A_{\|}|^2$ & $|A_{\perp}|^2$&$A^{dir}_{CP}$ \\
\hline \hline & 0.50&0.41 & 0.38 & 0.28 & 0.34&-0.63 \\ $B_s(\bar
B_s) \to \rho^0 \bar{K}^{\ast0}(K^{*0})$ & 0.55&0.34 & 0.41 & 0.27
& 0.32&-0.67
\\& 0.60&0.30 & 0.44 & 0.26 & 0.30&-0.70 \\ \hline
&0.50&0.56&0.33&0.31&0.36&0.56\\
$B_s(\bar B_s)
\to \omega \bar K^{\ast0}(K^{*0})$&0.55&0.47&0.35&0.30&0.35&0.58\\
&0.60&0.40&0.37&0.29&0.34&0.60\\
\hline
&0.50&16&0.92&0.04&0.04&0.10\\
$B_s(\bar B_s) \to \rho^\pm K^{*\mp}$&0.55&12&0.92&0.04&0.04&0.12\\
&0.60&10&0.92&0.04&0.04&0.13\\
\hline
\end{tabular}\end{table}

As we mentioned before, the decay $B_s \to \rho(\omega) K^*$ can
be used to determine the $B_s$ meson wave function parameter
$\omega_{B_s}$, or $\omega_{B_s}$ can influence our predictions of
$B_s \to \rho(\omega) K^*$ decay. So we show the results in
Table~1 according to 3 different values of $\omega_{B_s}$.
 From the table we can easily find out the averaged branching ratio
for $B_s(\bar B_s) \to \rho^\pm K^{*\mp}$ is much larger than the
other two, for $B_s(\bar B_s) \to \rho^\pm K^{*\mp}$ involve large
Wilson coefficient $C(2)+C(1)/3$ for the factorizable part while
the other two ($B_s(\bar B_s) \to \rho^0 \bar K^{*0}(K^{*0})$ and
$B_s(\bar B_s) \to \omega \bar K^{*0}(K^{*0})$) involve a much
smaller Wilson coefficient $C(1)+C(2)/3$ (color-suppressed) for
the factorizable part of the emission diagram. As a result the
first one is tree dominated and has a large branching ratio and
small direct CP asymmetry. While referring to the other two, the
contributions from penguin and tree diagrams are at the same order
($Z_H\simeq 0.5\sim1.5$), hence we can expect a large direct CP
asymmetry from eqs.(\ref{adircp}).

The polarization fraction difference of these channels are also
due to that the main contribution of each channel comes from
different topology. $B_s(\bar B_s) \to \rho^\pm K^{*\mp}$ is tree
dominated. The main contribution comes from the factorizable part
of the emission diagram, where transverse polarization amplitude
is suppressed by a factor $r_{\rho}^2 \simeq (0.77/5.37)^2$ (see
formulas in \cite{kstarkstar}), so the longitudinal polarization
dominates and contributes more than $90\%$ of the total branching
ratio. But in $B_s(\bar B_s) \to \rho^0 \bar K^{*0}(K^{*0})$ and
$B_s(\bar B_s) \to \omega \bar K^{*0}(K^{*0})$ decays, tree
emission (factorizable) diagram contribution is suppressed due to
the cancellation of Wilson coefficients $C_1+C_2/3$. The left
dominant contribution is the non-factorizable diagrams of tree
operators and penguin diagrams. both of these contributions
equally contribute to longitudinal and transverse polarizations.
The transverse polarization is not suppressed in those cases,
therefore numerically we get a small longitudinal fraction of
about 0.4. This similar situation is also found in $B\to \rho \rho
(\omega)$ decays \cite{rhorho}, which are related by SU(3)
symmetry to our $B_s \to\rho K^*$ decays.

\begin{figure}[tbh]
\begin{center}
\epsfxsize=3.0in\leavevmode\epsfbox{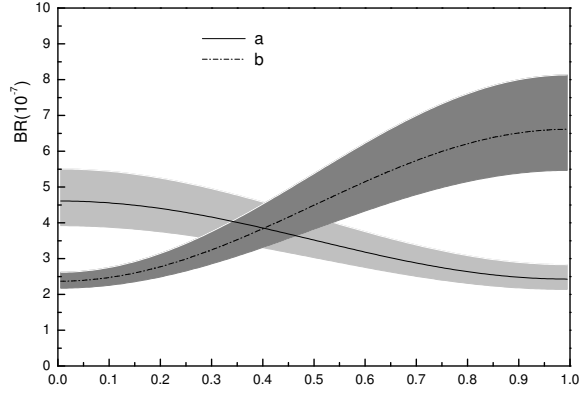}
\caption{{\protect\footnotesize Average branching ratios when
$\omega_{B_s}$ was set from 0.50   to 0.60: $(a)  B_s(\bar
B_s)\rightarrow \rho^0 \bar K^{*0}(K^{*0})$; (b) $ B_s(\bar
B_s)\rightarrow \omega \bar K^{*0}(K^{*0})$ }}
\end{center}
\end{figure}

    To extract the CP  violation parameters and dependence on CKM phase angle
$\alpha$ of these decays,  we rewrite the helicity amplitudes in
(\ref{m},\ref{barm}) as the functions of $\alpha$:
\begin{eqnarray}
{\cal M}_{H}^+&=&V_u^*T_H-V_t^{*}P_H \nonumber\\
&=&V_u^*T_H(1+Z_He^{i(\alpha+{\delta}_H)})\label{m++}\\
 {\cal
M}_{H}^-&=&V_uT_H-V_tP_H\nonumber\\
&=&V_uT_H(1+Z_He^{i(-\alpha+{\delta}_H)})\label{m--}
\end{eqnarray}where $Z_H=|V_t^*/V_u^*||P_H/T_H|$, and $\delta$ is the
relative strong phase between tree ($T$) and penguin ($P$)
diagrams. Here in PQCD approach, the strong phase comes from the
nonfactorizable diagrams and annihilation diagrams. This is
different from Beneke-Buchalla-Neubert-Sachrajda \cite{bbns}
approach. In that approach, annihilation diagrams are not taken
into account, strong phases mainly come from the so-called
Bander-Silverman-Soni mechanism \cite{bss}. As shown in
\cite{keum}, these effects are in fact next-to-leading-order
($\alpha_s$ suppressed) elements and can be neglected in PQCD
approach. We give the averaged branching ratios of $B_s(\bar B_s)
\to \rho^0 K^{*0}(\bar K^{*0})$ and $B_s(\bar B_s) \to \omega
K^{*0}(\bar K^{*0})$ as a function of $\alpha$ in Fig.3, and the
averaged branching ratios of $B^0(\bar B^0)\to \rho^{\pm}K^{*\mp}$
in Fig.4.

\begin{figure}
\begin{center}
\epsfxsize=3.0in\leavevmode\epsfbox{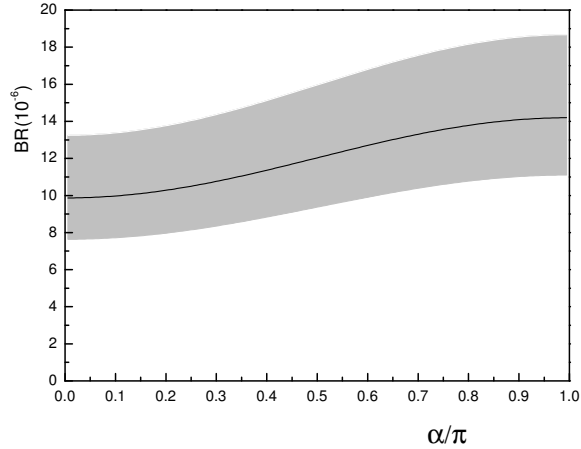}\end{center}
\vspace{-0.5cm}
\caption{{\protect\footnotesize Average branching
ratios for $B_s(\bar B_s)\rightarrow \rho^\pm K^{*\mp}$as a
function of $\alpha$ when $\omega_{B_s}$ was set from  0.50 to
0.60 }}
\end{figure}

Using Eqs.(\ref{m++},\ref{m--}), the direct CP violating parameter
is
\begin{eqnarray}
A^{dir}_{CP}&=&\frac{{|M|}^2-{|\bar M|}^2}{{|M|}^2+{|\bar
M|}^2}\nonumber
\\
&=&\frac{-2sin\alpha\left(T_L^2sin\delta_L+2T_N^2sin\delta_N+2T_T^2sin\delta_T
\right)}{T_L^2(1+Z_L^2+2Z_Lcos\alpha
cos\delta_L)+2\sum_{i=N,T}T_i^2\left(1+Z_i^2+2Z_icos\alpha
cos\delta_i\right) }.\label{adircp}
\end{eqnarray}
 Notice the CP asymmetry for these channels are sensitive to
CKM angle $\alpha$, we show the direct CP asymmetry as a function
of $\alpha$ in Fig.5. It is easy to see that the $B_s \to \rho^0
K^{*0}$ and $\omega K^{*0}$ have large direct CP asymmetries up to
50\%, with a relative minus sign. On the other hand, the $B_s \to
\rho ^\pm K^{*\mp}$ decay has small direct CP asymmetry due to
only one large tree contribution in this decay. The uncertainty
shown at this table is only from the $B_s$ meson wave function
parameter dependence. In fact, since CP asymmetry is sensitive to
many parameters, the line should be more broadened by
uncertainties. The mixing induced CP asymmetry is complicated and
requires angular distribution study, similar study may be found in
\cite{azimuthal}.

\begin{figure}[tbh]
\begin{center}\epsfxsize=3.0in\leavevmode\epsfbox{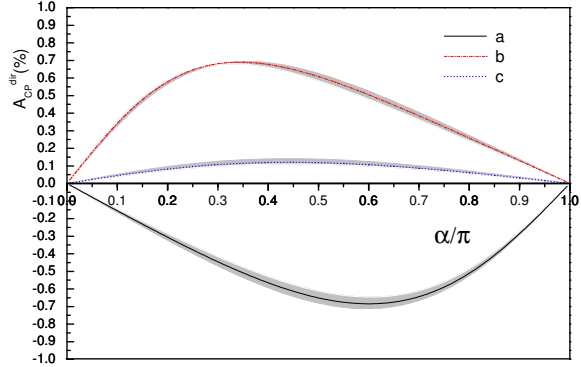}
\caption{Direct CP asymmetry $A_{CP}^{dir}$ as a function of
$\alpha$: (a) $\ B_s(\bar B_s)\rightarrow \rho^0  \bar
K^{*0}(K^{*0})$ (lower line);
 (b) $
B_s(\bar B_s)\rightarrow \omega  \bar K^{*0}(K^{*0})$ (upper
line); (c) $ B_s(\bar B_s)\rightarrow \rho^\pm K^{*\mp}$ (middle
line).} \end{center}\end{figure}

At last, if we compare our predictions with those of naive
factorization \cite{bsrhokfa}
\begin{eqnarray}
BR(B_s(\bar
B_s) \to \rho^0 \bar{K}^{\ast0}(K^{*0}))=5.5\times 10^{-7},\\
BR(B_s(\bar B_s) \to \omega \bar{K}^{*0}(K^{*0}))=6.0\times 10^{-7},\\
BR(B_s(\bar B_s) \to \rho^\pm K^{*\mp})=1.7\times 10^{-5},
\end{eqnarray}
 and ones of QCDF \cite{yangyd}
\begin{eqnarray}
BR(B_s(\bar
B_s) \to \rho^0 \bar{K}^{\ast0}(K^{*0}))=5.3\times 10^{-7},\\
BR(B_s(\bar B_s) \to \omega \bar{K}^{*0}(K^{*0}))=3.1\times 10^{-7},\\
BR(B_s(\bar B_s) \to \rho^\pm K^{*\mp})=1.8\times 10^{-5}.
\end{eqnarray}
We can see that they are consistent.  It should be noticed that
the branching ratios in FA and QCDF strongly depend on form
factors. While in PQCD, the branching ratios and form factors
depend on wave functions, especially the $B_s$ meson wave
function. Nowadays, very few $B_s$ meson decays have been
measured, so we can only give rough constraints on the parameters
 from other channel and permit large errors. More experimental data
can help to constrain the form factors and wave functions, then we
can give more precise predictions and the different methods can be
tested by the experiments. Although similar results are got by
different methods for branching ratios, the polarization fractions
are quite different. The QCDF and naive factorization give only
several percent transverse polarization for all three decay modes
\cite{yangyd}, while Table 1 shows large transverse contribution
for $B_s \to \rho^0 (\omega) K^{*0}$ decays in our PQCD approach.
The direct CP asymmetry are not given in ref.\cite{yangyd}, but
they probably also differ from PQCD approach as it happened in
$B\to\pi\pi $ and $K\pi$ case \cite{cp}.

The numerical results shown here are only leading order ones. For
the tree dominated channel $B_s \to \rho^{\pm} K^{*\mp}$, the
leading order diagrams should give the main contribution.  But for
the other two decays, with a branching ratio as small as
$10^{-7}$, the next-to-leading order and power suppressed
contributions should not be negligible, the results may suffer
 from large corrections when the next to leading order corrections
are included \cite{next}.

Current experiments \cite{particle} only give the upper limit for
the decay
\begin{eqnarray}
BR(B_s \to \rho^0\bar{K}^{\ast0})<7.6\times 10^{-4}.
\end{eqnarray}
More data are needed to test our calculations.

\section{Summary}

\hspace*{\parindent}In this paper we calculate the branching
ratios, polarization fraction and CP asymmetries of $B_s \to
\rho(\omega)K^*$ modes using PQCD theorem in SM. We perform all
leading order diagrams to next to leading twist wave functions. We
also study the dependence of their averaged branching ratios and
the CP asymmetry on the CKM angle $\alpha$. At last we compare our
predictions with   values from other approaches.

\section*{Acknowledgments}

\hspace*{\parindent}This work is partly supported by the National
Science Foundation of China under Grant (No.90103013, 10475085 and
10135060). We thank G.-L. Song for reading our manuscript and
giving us many helpful suggestions. We also thank J-F Cheng, H-n
Li, Y. Li, and X-Q Yu for helpful discussions. We thank the
Institute for Nuclear Theory at the University of Washington for
its hospitality and the Department of Energy for partial support
during the completion of this work.

\begin{appendix}

\section{wave function}
\hspace*{\parindent}

For longitudinal polarized $K^*$ meson, the wave function is
written as
\begin{eqnarray}
\frac{1}{\sqrt{2N_c}}[M_{K^*}\not\!\epsilon_{2L}\phi_{K^*}(x)+\not\!\epsilon_{2L}
\not\!P_2\phi_{K^*}^t(x)+M_{\phi}I\phi^s_{K^*}(x)],
\end{eqnarray}
and  the wave function for transverse polarized $K^*$ meson reads
\begin{eqnarray}
\frac{1}{\sqrt{2N_c}}[M_{K^*}\not\!\epsilon_{2T}\phi^v_{K^*}(x)+\not\!\epsilon_{2T}
\not\!P_2\phi_{K^*}^T(x)+\frac{M_{K^*}}{P_2\cdot
n_-}i\epsilon_{\mu\nu\rho\sigma}\gamma_5\gamma^{\mu}\epsilon_{2T}^{\nu}P_2^{\rho}n_-^{\sigma}\phi^a_{K^*}(x)].
\end{eqnarray}

The $K^*$ meson distribution amplitudes up to twist-3 are given by
ref.\cite{bbkt} with QCD sum rules.
\begin{eqnarray}
&&\phi_{K^*}(x)=\frac{3f_{K^*}}{\sqrt{2N_c}}x(1-x)[1+0.57(1-2x)+0.07C_2^{3/2}(1-2x)],\\
&&\phi^t_{K^*}(x)=\frac{f_{K^*}^T}{2\sqrt{2N_c}}\left\{0.3(1-2x)[3(1-2x)^2+10(1-2x)-1]
+1.68C_4^{1/2}(1-2x)\right.\nonumber\\
&&
\left.\ \ \ \ \ \ \ \ \ \ \ \ +0.06(1-2x)^2[5(1-2x)^2-3]+0.36\left\{1-2(1-2x)[1+\ln(1-x)]\right\}\right\},\\
&&\phi^s_{K^*}(x)=\frac{f_{K^*}^T}{2\sqrt{2N_c}}\left\{3(1-2x)\left[1+0..2(1-2x)+0.6(10x^2-10x+1)\right]
\right.\nonumber\\
&& \left.\ \ \ \ \ \ \ \ \ \ \ \
-0.12x(1-x)+0.36[1-6x-2\ln{(1-x)}]\right\},\\
&&\phi^T_{K^*}(x)=\frac{3f_{K^*}^T}{\sqrt{2N_c}}x(1-x)[1+0.6(1-2x)+0.04C_2^{3/2}(1-2x)
],\\
&&\phi^v_{K^*}(x)=\frac{f_{K^*}^T}{2\sqrt{2N_c}}\left\{\frac{3}{4}[1+(1-2x)^2+0.44(1-2x)^3]
\right.\nonumber\\
&& \left.\ \ \ \ \ \ \ \ \ \ \ \ +0.4C_2^{1/2}(1-2x)
+0.88C_4^{1/2}(1-2x)+0.48[2x+\ln(1-x)]
\right\},\\
&&\phi^a_{K^*}(x)=\frac{f_{K^*}^T}{4\sqrt{2N_c}}\left\{3(1-2x)[1+0.19(1-2x)+0.81(10x^2-10x+1)]
\right.\nonumber\\
&& \left.\ \ \ \ \ \ \ \ \ \ \ \
-1.14x(1-x)+0.48[1-6x-2\ln(1-x)]\right\},
\end{eqnarray}
where the Gegenbauer polynomials are
\begin{eqnarray}
&&C_2^{\frac{1}{2}}(\xi)=\frac{1}{2}(3\xi^2-1),\\
&&C_4^{\frac{1}{2}}(\xi)=\frac{1}{8}(35\xi^4-30\xi^2+3),\\
&&C_2^{\frac{3}{2}}(\xi)=\frac{3}{2}(5\xi^2-1).
\end{eqnarray}

For $\rho$ and $\omega$ meson, we employ
$\rho^0=\frac{1}{\sqrt2}(u\bar u-d\bar d)$ and
$\omega=\frac{1}{\sqrt2}(u\bar u+d\bar d)$. Their Lorentz
structures are similar to $K^*$ meson, the  distribution
amplitudes are the same for $\rho$ and $\omega$ and given as
\cite{bbkt}:
\begin{eqnarray}
\phi_\rho(x)&=&\frac{3f_\rho}{\sqrt{2N_c}} x(1-x)\left[1+
0.18C_2^{3/2}(1-2x)\right]\;,
\label{pwr}\\
\phi_{\rho}^t(x)&=&\frac{f^T_{\rho}}{2\sqrt{2N_c}}
\left\{3(1-2x)^2+0.3(1-2x)^2[5(1-2x)^2-3]\right.
\nonumber \\
& &\left.+0.21[3-30(1-2x)^2+35(1-2x)^4]\right\}\;,
\label{pwt}\\
\phi_{\rho}^s(x) &=&\frac{3f_\rho^T}{2\sqrt{2N_{c}}}
(1-2x)\left[1+0.76(10x^2-10x+1)\right]\;,
\label{pws}\\
\phi_\rho^T(x)&=&\frac{3f_\rho^T}{\sqrt{2N_c}} x(1-x)\left[1+
0.2C_2^{3/2}(1-2x)\right]\;,
\label{pwft}\\
\phi_{\rho}^v(x)&=&\frac{f_{\rho}}{2\sqrt{2N_c}}
\bigg\{\frac{3}{4}[1+(1-2x)^2]+0.24[3(1-2x)^2-1]
\nonumber \\
& &+0.12[3-30(1-2x)^2+35(1-2x)^4]\bigg\}\;,
\label{pwv}\\
\phi_{\rho}^a(x) &=&\frac{3f_\rho}{4\sqrt{2N_{c}}}
(1-2x)\left[1+0.93(10x^2-10x+1)\right]\;. \label{pwa}
\end{eqnarray}

\end{appendix}

\end{document}